# The Plastic Flow of Solid $^4$He through a Porous Membrane

A. Lisunov, V. Maidanov, A. Neoneta, V. Rubanskyi, S. Rubets, E. Rudavskii, and V. Zhuchkov

*B. Verkin Institute for Low Temperature Physics and Engineering of the National Academy of Sciences of Ukraine, Kharkov, Ukraine*

rudavskii@ilt.kharkov.ua

**Keywords:** Solid Helium, Plastic Flow, Vacancies

### Abstract

The flow velocity of solid $^4$He through a porous membrane frozen into a crystal has been measured in the temperature interval 0.1 – 1.8 K. A flat capacitor consisting of a metalized plastic porous membrane and a bulk electrode is applied and the gap in the capacitor is filled with examined helium. The flow of helium through the membrane pores is caused by a d.c. voltage applied to the capacitor plates. Above T~1 K the velocity of solid $^4$He flow decreases with lowering temperature following the Arrhenius law with the activation energy of the process closed to that of vacancies. At low temperatures the velocity is practically independent of temperature, which suggests a transition in $^4$He from the classical thermally activated vacancy-related flow to the quantum plastic flow.

**PACS**: 67.80.-s, 67.25. du, 67.30. ef

## 1. Introduction

The anomalous behavior of solid $^4$He in the temperature region below ~300 mK is still open to arguments and therefore attracts huge interest of both experimenters and theorists (e.g., see surveys [1,2]). One of the possible scenarios considers a change in the plasticity of solid helium.

Plastic flow in solid helium was first detected by Suzuki [3] while he was watching the motion of a small ball frozen into a crystal. The observed phenomenon was interpreted within a dislocation model. The dislocation nature of plastic deformation in solid helium was supported in the experiments by Tsymbalenko [4] who investigated the motion of a plate in solid helium. Later a series of experiments [5-7] was made to investigate the flow of solid helium through a porous membrane. The results obtained were explained involving the vacancy mechanism of mass transfer through a membrane. Search for a pressure-driven flow of solid helium was made by J.Day and J.Beamish [8] using a piezoelectric diaphragm. They found that the average flow velocity at low temperatures was less than $1.2 \cdot 10^{-14}$ m/s.

New interesting features of plasticity of solid $^4$He were revealed while measuring its shear modulus [9-12]. It was found that the shear modulus increased considerably as the temperature lowered below ~200 mK which points to stiffening the crystal and a loss of plasticity. The effect was explained by a change of the motion of dislocations: they move freely at high temperature and thus ensure high plasticity; below the temperature mentioned they are pinned by $^3$He impurities and lose their mobility. The experiments on ultrapure $^4$He single crystals [12] demonstrated giant plasticity of the crystal which was due to the free dislocation glide in the basal plane of the hcp structure. The change in the plastic properties of solid helium were interpreted theoretically [13, 14] within the dislocation model.

The goal of this study is to investigate the plastic flow of solid helium through a porous membrane in a wide range of temperatures including the regions of classical thermally activated and quantum plasticity.

## 2. Experimental technique

The motion of a porous membrane frozen into solid helium was investigated using the capacitive method similarly to [5-7]. The measuring cell is shown schematically in Fig. 1. Porous membrane 1 and immovable bulk electrode 2 form a flat capacitor filled with sample 3 of solid helium. As d.c. voltage $U$ is applied to the capacitor plates, it generates a force $F$ moving the membrane in solid helium. The membrane was made of aluminized lavsan of 10 μm thickness with an irregular system of 6 – 8 μm holes and 18% porosity.

The changes in the capacitance of the capacitor (membrane 1 plus electrode 2) in course of the measurement were registered using BR-2827 capacitance bridge. Separating capacitors 5 protected the capacitance bridge from the applied voltage. The measuring circuit also included high-value resistors 8 which could restrict the current in case of the membrane breakdown.

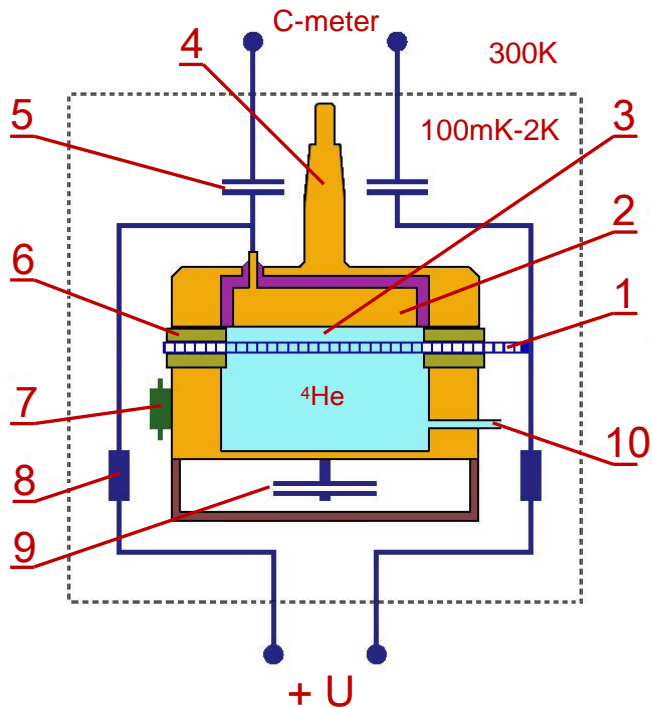

Fig. 1. The schematic view of the measuring cell. 1 – metalized porous membrane, 2 – the stationary electrode, 3 – solid $^4$He sample, 4 – cooling link, 5 – separating capacitors, 6 – kapton spacer, 7 – RuO$_2$ resistance thermometer, 8 – restricting resistors, 9 – capacitance pressure gauge, 10 – filling capillary.

The measuring cell was cooled from the mixture chamber through cooling link 4 and its temperature was measured with RuO$_2$ resistance thermometer 7 calibrated against a melting curve of $^3$He thermometer. The cell was also equipped with capacitive pressure gauge 9 to register the pressure of solid helium *in situ*.

Prior to growing samples of solid $^4$He, the membrane was calibrated in liquid helium. The capacitance was measured at zero potential and the applied voltages used in the experiment (0 – 400 V). The calibration made possible to find the equilibrium non-deformed position of the membrane and to control it in the course of the experiment on solid helium. Note that the displacement of the membrane in solid helium is so small that the influence of the membrane elasticity becomes negligible.

The solid $^4$He samples (concentration of the $^3$He impurities was about 1ppm) were grown by the blocking capillary technique. The sample starts to grow at the top between the fixed electrode and the membrane. In the process of growth it was seen as the membrane is displaced from its equilibrium position (the capacitance is reduced). To return the membrane in its equilibrium state, the crystal is partially re-melted. This was made several times until the membrane occupied the position of equilibrium, which was known from the calibration measurements in superfluid helium. The equilibrium state of the crystal was controlled by a capacitive pressure gauge 9 (Fig.

1) according to stopping the change in pressure. The procedure described above is equivalent to annealing of the crystal, so only good quality crystals are investigated.

The experiments were made at a constant molar volume 20.8 cm³/mol. We measured the change in the capacitance $\Delta C$ of the capacitor filled with solid $^4$He during the time $\Delta t$ after applying d.c. voltage $U$. The clearance $l$ between the membrane and the stationary electrode of the flat capacitor varied with time as $l(t) = (\varepsilon\, S)/4\pi C(t)$. The value of the membrane velocity is

$$V = \frac{\Delta l}{\Delta t} = \frac{\varepsilon\, S}{4\pi C^2} \frac{\Delta C}{\Delta t}, \qquad (1)$$

where $S$ is the membrane area. The experiments performed have provided important information about the temperature dependence of velocity $V$.

### 3. Results and discussion

The time dependence of the capacitance after applying the voltage $U$ to the capacitor plates is illustrated in Fig. 2. The $U$ value determines the acting force $F=(CU^2)/2l$ and the pressure on the solid $^4$He sample

$$P = \frac{2\pi C^2 U^2}{\varepsilon S^2}. \qquad (2)$$

After applying the voltage $U$, the capacitance $C$ varies with time following a linear law (Fig. 2) and the slope of the corresponding linear dependence of the capacitance increased with growing $U$.

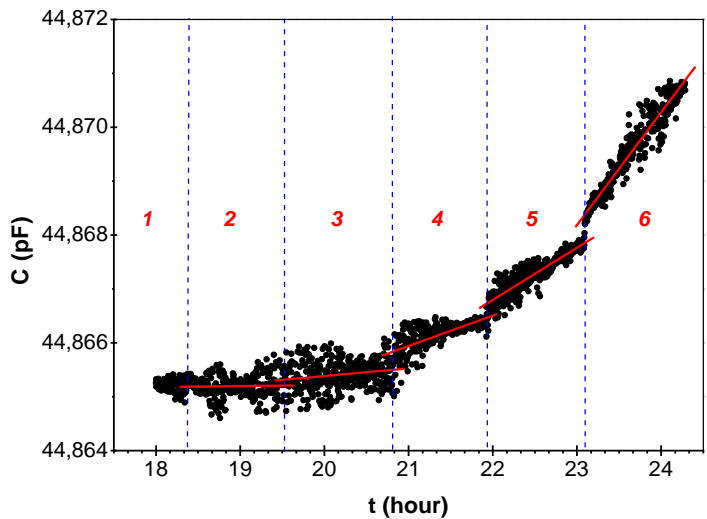

Fig. 2. The kinetics of the capacitance $C$ of the capacitor at different $U$-values: U=0 (1), 50V (2), 100V (3), 200V (4), 300V (5), 400V (6). The temperature T=1.52K.

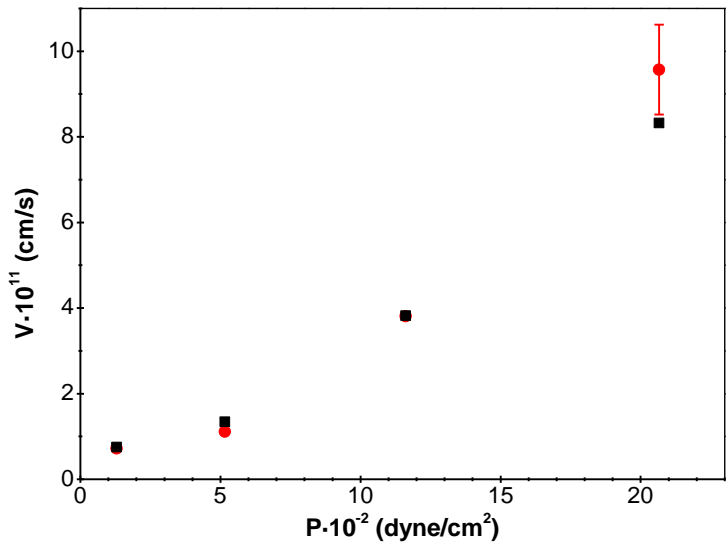

Fig. 3. The membrane velocity as a function of the mechanical stress in solid $^4$He. T=1.52 K: ■ – on increasing the applied voltage, ● – on decreasing the applied voltage.

Using the above dependence and Eqs. (1) and (2) we obtain the velocity *V* for each mechanical stress *P* (see Fig. 3).

The temperature dependence of the membrane velocity *V* corresponding to the plastic flow rate of solid helium is shown in Fig 4. It is seen that in the high-temperature range the velocity *V* decreases rather fast with the lowering temperature. Below ~ 1K the velocity is practically independent of temperature. The absolute value of the velocity is less than $10^{-11}$ cm/s under the voltage *U* = 400V. This value coincides with an inaccuracy in finding of *V* which is determined by the inaccuracy of *ΔC / Δt* ($0.7·10^{-7}$ pF/s).

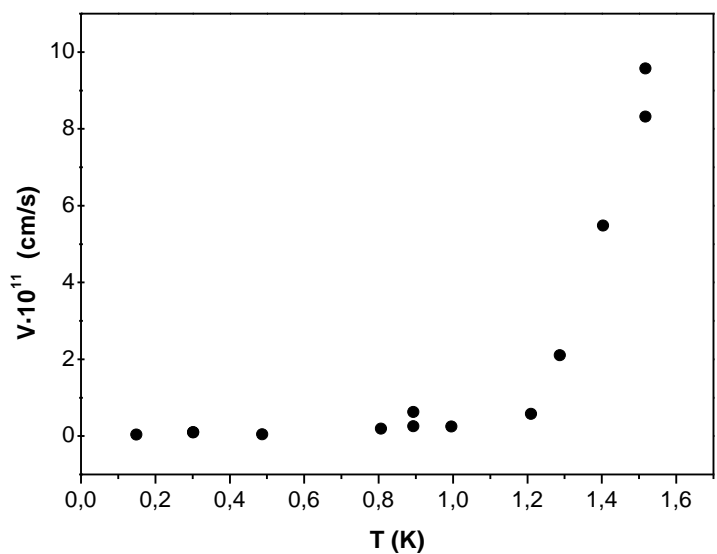

Fig. 4. The temperature dependence of the membrane motion velocity in solid helium at the voltage *U* = 400 V.

Thus, the experiments revealed that under a constant external load solid he-lium "flows" through the holes in the membrane, i.e., plastic flow of the crystal is observed. Since the plastic deformation of the crystal is related to an irreversible change in the relative positions of the atoms in the lattice and this change most easily occurs in areas of distorted crystal, the plastic effects commonly associated with a variety of structural defects – vacancies, dislocations, etc.

To identify potential mechanisms of plastic solid helium flow, the data obtained are represented in logarithmic scale depending on the inverse temperature. It allows to determine the activation energy of the process. These data are shown in Fig. 5 which shows that above 1 K temperature dependence velocity $V$ can be approximated by an exponential dependence (Arrhenius law):

$$V = V_o \, exp \, (-E_a / T), \qquad (3)$$

where the pre-exponential factor $V_o$ = $1.1 \cdot 10^{-7}$ cm/s and the exponent has the meaning of the activation energy $E_a = 11.1 \pm 1.5$ K.

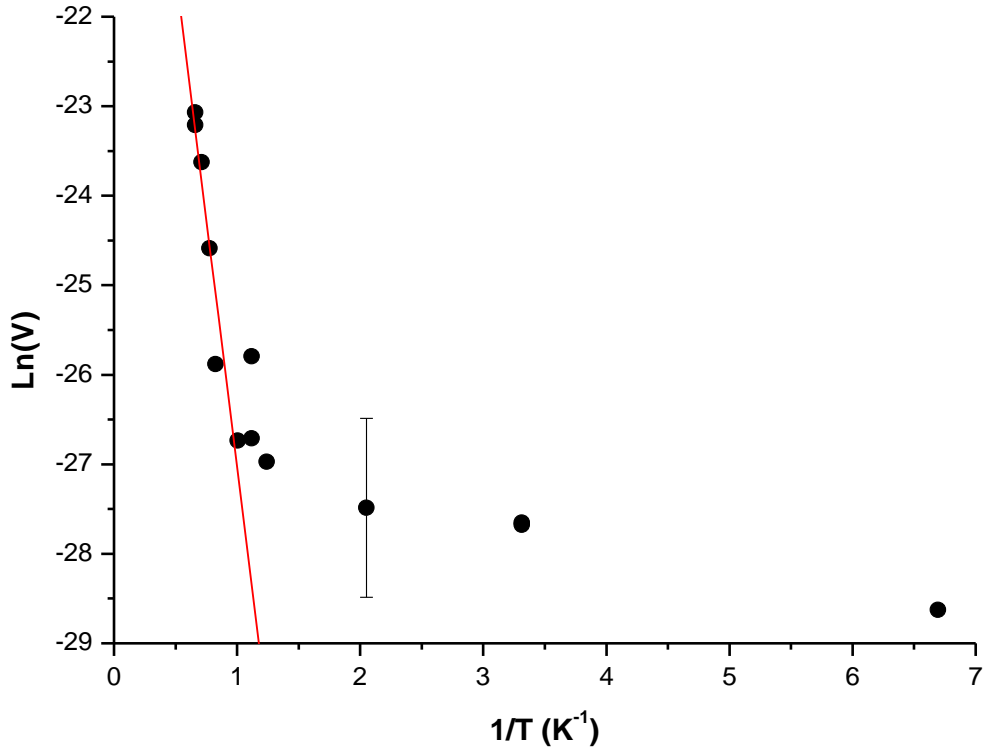

Fig. 5. The dependence of the velocity of $^4$He flow through a porous membrane upon inverse temperature. Solid line is the calculation by Eqs. (3).

This means that the plastic flow in the specified temperature region is a thermally activated process with the activation energy $E_a$. The resulting value of $E_a$ is in agreement with the activation energy of vacancies $E_v$, the value of which was obtained earlier on the basis of analysis of numerous experimental data available [15] for the hcp phase of solid $^4$He. Note that under experimental conditions the contribution to the plastic flow gives both thermal and nonequilibrium vacancies.

Thus, in this temperature range vacancy diffusion mechanism of flow of solid helium is realized, that is consistent with the results of [5-7]. The structure of the crystal lattice is not disturbed, and inelastic change in the shape of the crystal is due to directional flow of vacancies. The reason for this is the stress in the crystal produced by the deformation of the membrane.

In this case, the flow velocity $V$ is proportional to the flux of atoms $j$ (or vacancies in the opposite direction) $V \sim j$, and $j$ in turn is proportional to the self-diffusion coefficient of helium atoms $D_S$. The value of $D_S$ is equal to the product of the vacancy diffusion coefficient $D_v$ to their concentration $x_v$. As a result, the helium flow rate is proportional to the concentration of vacancies $x_v \sim exp\ (-E_v / T)$.

With regard to sources and drains of vacancies which are necessary to realize the diffusion-vacancy crystal flow, then they may be the outer surfaces of the crystal, grain boundaries and dislocations occurring during crystal growth and loading.

In the temperature range below ~ 1 K, as can be seen from Figs. 4 and 5, the character of the temperature dependence of the flow rate varies essentially. In this case one should expect a significant influence of quantum effects on the plastic flow of solid helium. The fact that below ~ 1 K the flow rate is almost independent of temperature (within the measurement bars) may indicate a transition from classical thermally activated to quantum plastic flow of the crystal. The question of the possible mechanisms of such a flow, and the role of vacancies and dislocations in the process requires separate consideration.

4. Conclusions

The described experiments on the flow of solid helium through a porous membrane showed that two areas can be distinguished in the temperature dependence of the flow velocity – high temperature and low temperature region. At temperatures above ~ 1 K the velocity reduces rapidly with decreasing temperature, which may be described in the model of thermally activated diffusion flow of vacancies. At low temperatures, the flow velocity is almost independent of temperature, indicating the quantum character of the plastic flow of solid helium.

Note, that the highest electric voltage to the capacitor plates attainable in the experiment was up to 400 V, which corresponds to the mechanical stress $2.1\cdot10^3$ dyne/cm$^2$. Apparently, at low enough temperature, such a stress was insufficient to observe plastic flow of the solid helium through the holes in the membrane. The corresponding experiments will be continued in the area of higher stresses.

**Acknowledgements**

The authors are grateful to V.D. Natsik and S.N. Smirnov for helpful discussions and to N.V. Isaev, A.S. Rybalko, A.P. Birchenko and N.P. Mikhin for help in preparation of the experiment.